\documentstyle[12pt]{article}
\begin{document}

\title{Numerical Computation of the Effective Potential and Renormalization}
\author{G. Palma and L. Vergara \\
%EndAName
Departamento de F\'\i sica, Universidad de Santiago de Chile\\
Casilla 307, Correo 2, Santiago de Chile.}
\date{}
\maketitle

\begin{abstract}
We present a novel way to compute the one-loop ring-improved effective
potential numerically, which avoids the spurious appearence of complex
expressions and at the same time is free from the renormalization
ambiguities of the self-consistent approaches, based on the direct
application of Schwinger-Dyson type equations to the masses.
\end{abstract}

\section{Introduction}

In the study of the nature of the electroweak phase transition in the early
Universe the effective potential plays an essential role. Since the early
studies it was realized that the computation of the effective potential was
plagued with difficulties which are mainly related with the bad infrared
behavior of the theory near the critical temperature \cite{DOLAN}. In order
to avoid the appearance of imaginary terms in the effective potential it was
then necessary to improve its perturbative computation by resumming an
infinite subset of the leading infrared divergent diagrams, called the ring
diagrams. An important improvement was made by the use of self-consistent
methods based on Schwinger-Dyson equations for the masses. They can be
obtained through a modification of the mass term in the original Lagrangian;
a mass parameter is introduced in the beginning and later determined by
consistency conditions, the gap equations (see e.g. \cite{GROSS},\cite{BUCH}%
).

In spite of the considerable progress made in this direction, there still
remain some ambiguities in the calculation of the effective potential at
finite temperature which are mainly related with the correct way to perform
the resummation of the infrared divergent diagrams. Also when using the kind
of methods based on gap equations it will always appear temperature
dependent ultraviolet divergencies, which implies that in order to
renormalize the theory one should include temperature dependent
counterterms. Although these counterterms have the same structure as those
of the $T=0$ theory, its physical origin is very unclear.

In this letter we show a novel way of computing the one-loop ring-improved
effective potential numerically which avoid the above mentioned problems
related with renormalization. It is enough to consider the $\lambda \phi ^4$
-theory to explain the method, but for definitness the final calculation is
carried out within the Standard Model.

\section{The self-consistent way}

We begin by reviewing (in the $\lambda \phi ^4$-theory) the usual way one
performs the improvement of the one-loop effective potential at finite
temperature. We add and substract in the original lagrangian the induced
thermal mass $M^2=\Pi _T(\omega _n,p)$, where $\omega _n=2\pi nT$, $n=0,\pm
1,...$ are the Matsubara frequencies. The result of this procedure is an
improved perturbation theory in which the original propagator is replaced by
the full propagator. The polarization tensor $\Pi _T(\omega _n,p)$ can be
obtained consistently within perturbation theory. Due to the fact that it is
the infrared behavior the relevant aspect near the phase transition, it is
only necessary to compute $\Pi _T$ in the infrared limit $\omega _n=0$, $%
p\rightarrow 0$, $\Pi _T(0)$. This procedure amounts to a summation of an
infinite set of diagrams with the worst infrared behavior, called the ring
diagrams, and which are generated by including polarization tensor
insertions in the infrared limit in one loop diagrams. It is not necessary
to do that for fermions since in this case the Matsubara frequencies are
always different from zero. Therefore, the effective potential reads

\begin{equation}
U_{eff,T}(\phi)=U_{cl}(\phi)+\frac {T}{2}\sum_n\int \frac{d^3k}{(2\pi)^3}
\ln \left[ \omega_n^2+{\bf \vec k}^2+m_{pl}^2(\phi)\right],  \nonumber
\end{equation}
where $m_{pl}^2(\phi)=m^2(\phi)+\Pi_T(0)$ is the plasma mass ($%
m^2(\phi)=-\nu ^2+\lambda \phi^2/2)$. The sum over Matsubara frequencies can
be performed and one obtains

\begin{equation}  \label{prim}
U_{eff,T}(\phi )=U_{cl}(\phi )+\frac 12\int \frac{d^3k}{(2\pi )^3}\sqrt{{\bf %
\vec k}^2+m_{pl}^2(\phi )}+\frac T{2\pi ^2}\int dk{\bf \vec k}^2\ln \left[
1-e^{-\sqrt{{\bf \vec k}^2+m_{pl}^2(\phi )}/T}\right]  \nonumber
\end{equation}
It is clear that the second term on the r.hs. of this expression contains
temperature dependent ultraviolet divergent contributions. Therefore
temperature dependent counterterms are needed in order to renormalize this
expression. In spite of the similar structure of these counterterms with
those of the $T=0$ limit, its physical origin is not clear and represents a
lack of consistency of the procedure used \cite{Mack}. In fact the
renormalization problem is related with the local behavior of the theory and
should not be influenced by finite temperature effects, which concerns long
distance physics.

Nevertheless if one accepts to work with temperature dependent counterterms
one can formally derive an expression for the effective potential. To find
explicit expressions see ref. \cite{CARR}.

\section{Our proposal}

In this section we show a novel way of computing the one-loop ring-improved
effective potential which avoids the above mentioned problems related with
renormalization.

In this approach we will consider the improvement of the effective potential
by the summation of rings diagrams, which in this case are generated by
including polarization tensor insertions in the infrared limit, in one-loop
diagrams with {\it zero Matsubara frequencies}. To explain the method we
start with the expression for the one-loop ring-improved effective potential

\begin{equation}  \label{pott}
\hat U_{eff,T}(\phi )=-\frac 12\nu ^2\phi ^2+\frac \lambda {4!}\phi ^4+\frac
T2\int \frac{d^3k}{(2\pi )^3}\ln \left[ {\bf \vec k}^2+m_{pl}^2(\phi
)\right] ,  \nonumber
\end{equation}
where the third term on the r.h.s can be separated into the one-loop

\begin{equation}  \label{unot}
\hat U_{eff,T}^{(1)}(\phi )=\frac T2\int \frac{d^3k}{(2\pi )^3}\ln \left[
{\bf \vec k}^2+m^2(\phi )\right]  \nonumber
\end{equation}
and the ring

\begin{equation}  \label{ringts}
\hat U_{eff,T}^{ring}(\phi )=-\frac T{12\pi }\left( \left[ m^2(\phi )+\Pi
_T(0)\right] ^{3/2}-m^3(\phi )\right)
\end{equation}
\noindent contributions.

At first sight, by looking at eqs. (\ref{unot}) and (\ref{ringts}) one
should be tempted to say that there is a difficulty in using the above
equation for the effective potential, because for small enough values of the
classical field $\phi $ the scalar mass squared $m^2(\phi )$ is negative and
therefore the potential itself would be a complex expression in that range.
Of course this is not so because although both $\hat U_{eff,T}^{(1)}(\phi )$
and $\hat U_{eff,T}^{ring}(\phi )$ become complex for small enough values of
$\phi $, their sum, given by equation (\ref{pott}), is real provided $%
m_{pl}^2(\phi )>0$. This is not new and people have used this fact to
evaluate the effective potential in the high-temperature expansion. In this
case the ''dangerous'' terms, i.e. those which become imaginary if $m^2(\phi
)<0$ cancel each other and everything is safe.

But what happens if we want to evaluate the one-loop ring-improved effective
potential in an exact way, that is, numerically? In this case a
calculational problem arises because the scalar mass squared are negative
for finite values of $\phi $. The usual way out of this problem is to carry
out the self consistent method outlined in the previous section which, as we
have seen, from first principles is not correct, because it mixes the
infrared and ultraviolet regimes.

The method we propose is based on eq. (\ref{pott}) and avoids all the above
mentioned difficulties. The important point to remember is that $\hat
U_{eff,T}(\phi )$ is real provided $m_{pl}^2(\phi )>0$.

The one-loop contribution to the effective potential $\hat U_{eff,T}(\phi)$
is

\begin{eqnarray}  \label{pott2}
\hat U^{(1)}_{eff,T}(\phi )&=&\frac {1}{64\pi ^2}\left\{ m^4(\phi )\left(
\ln \left[ \frac{m^2(\phi )}{m^2(\phi_0)}\right] -\frac {3}{2}\right)
+2m^2(\phi ) m^2(\phi _0)\right\}  \nonumber \\
&& +\frac{T^4}{2\pi ^2}\int_0^\infty dxx^2\ln \left[ 1-e^{-\sqrt{%
x^2+m^2(\phi )/T^2}}\right] .
\end{eqnarray}

\noindent The renormalization is performed here at zero temperature which in
turns implies that the required counterterms are temperature-independent. As
renormalization prescription we require that both the classical values of
the Higgs boson mass $m^2=2\nu ^2$ and the Higgs field at the minimum of the
effective potential at $T=0$, $\phi_0=\sqrt{6\nu^2/\lambda }$, to be
preserved even after quantum corrections are taken into account. In
opposition to what happens in the self-consistent approach, the
renormalization procedure is free of any theoretical inconsistency.

If $m^2(\phi )>0$ everything is standard and no difficulties appear in the
calculation of eq. (\ref{pott2}) . Therefore let us consider the case when $%
m^2(\phi )<0$ and define $\widehat{m}^2(\phi )\equiv -m^2(\phi )>0$. Thus,
by separating the range of integration in the temperature dependent part we
have

\begin{eqnarray}
\hat U_{eff,T}^{(1)}(\phi )&=& \frac{T^4}{2\pi ^2}\int_0^{\widehat{m}(\phi
)/T}dxx^2\ln \left[ 1-e^{-i \sqrt{\widehat{m}^2(\phi )/T^2-x^2}}\right]
\nonumber \\
&& +\frac{T^4} {2\pi ^2} \int_{\widehat{m}(\phi )/T}^\infty dxx^2\ln \left[
1-e^ {-\sqrt{x^2-\widehat{m}^2(\phi )/T^2}}\right] .
\end{eqnarray}

Clearly the second term is real, while the first one is complex and can be
explicited as follows. Notice that

\begin{eqnarray}
\ln \left[ 1-e^{-i \sqrt{\widehat{m}^2(\phi )/T^2-x^2}}\right]&=&-i\sqrt{%
\widehat{m}^2(\phi )/T^2-x^2}/2+\ln 2+i\frac {\pi}{2}  \nonumber \\
&& + \ln \left[ \sin \left( \frac {1}{2}\sqrt{\widehat{m}^2(\phi )/T^2-x^2}%
\right) \right] ,
\end{eqnarray}

\noindent where we have chosen the branch point where $\ln i=i\pi /2.$ Thus,

\begin{eqnarray}
\hat U_{eff,T}^{(1)}(\phi )&=&-i \frac{T^4}{2\pi ^2}\int_0^{\widehat{m}(\phi
)/T}dxx^2\left[ \sqrt{\widehat {m}^2(\phi )/T^2-x^2}-\pi \right]  \nonumber
\\
&& + \frac{T^4}{2\pi ^2}\int_0^{\widehat{m}(\phi)/T}dxx^2\ln \left[ 2\sin
\left( \frac 12\sqrt{\widehat{m}^2(\phi )/T^2-x^2}\right) \right]  \nonumber
\\
&& + \frac{T^4}{ 2\pi^2}\int_{\widehat{m}(\phi )/T}^\infty dxx^2\ln \left[
1-e^ {-\sqrt{x^2-\widehat{m}^2(\phi )/T^2}}\right] .
\end{eqnarray}

The imaginary part of $\hat U_{eff,T}^{(1)}(\phi)$ can be integrated exactly
and the result is

\[
Im\hat U_{eff,T}^{(1)}(\phi )=-\frac 1{64\pi ^2}\widehat{m}^4(\phi )+\frac
T{12\pi }\widehat{m}^3(\phi )
\]

\noindent Notice that the second term on the r.h.s. corresponds to one of
the terms appearing in the ring contribution in the scalar sector, see eq. (%
\ref{ringts}). Also notice that the presence of the extra term, which does
not appear in the ring contribution. Its effect is to cancel the imaginary
part of the zero temperature contribution coming from the logarithm
(remember that we are considering the case of negative mass squared).

This shows an exact cancellation of the spurious imaginary terms in the
effective potential, as it should be the case because for $m_{pl}^2(\phi)>0 $%
, the effective potential is real . Thus, it is clear that such a
cancellation does not depend on the order of the high-temperature expansion.

Therefore, the final expression for the effective potential, for the case $%
m^2(\phi )<0$, is

\begin{eqnarray}  \label{potefft}
\hat U_{eff,T}(\phi )&=&-\frac {1}{2}\nu ^2\phi ^2+\frac \lambda {4!} \phi
^4 + \frac{1}{64\pi ^2}\left \{ \widehat{m}^4(\phi)\left( \ln \left[ \frac{%
\widehat{m}^2(\phi)}{m^2(\phi_0)}\right] -\frac {3}{2}\right) -2\widehat{m}%
^2(\phi ) m^2(\phi_0)\right \}  \nonumber \\
&&+ \frac{T^4}{2\pi ^2}\int_0^{\widehat{m}(\phi)/T} dxx^2\ln \left[ 2\sin
\left( \frac {1}{2}\sqrt{\widehat{m}^2(\phi ) /T^2-x^2} \right) \right]
\nonumber \\
&& + \frac{T^4}{2\pi^2}\int_{\widehat{m}(\phi)/T}^\infty dxx^2\ln \left[
1-e^{-\sqrt{x^2-\widehat{m}^2(\phi )/T^2}}\right] .
\end{eqnarray}

\section{Numerical results}

As we said in the introduction, we present our numerical results in the
context of the Standard Model. The values of the SU(2) and U(1) couplings
constants are $g_2=0.647$ and $g_1=0.344$, respectively. Also, we take $\phi
_0=246$ GeV.

We implement the first (self-consistent) approach by following Carrington
\cite{CARR}, that is, by including the whole range of Matsubara frequencies
in the scalar sector and by considering only the contribution of the zero
Matsubara frequency in the gauge sector. In particular, this means that we
consider equations of the form (\ref{prim}) in order to evaluate the
temperature dependent part of the effective potential. More precisely, we
use the same set of equations that she used in order to evaluate the
effective potential\footnote{%
Our numerical results do not coincide with those of Carrigton when using the
same equations. Since we have checked the correctnes of our calculations, we
conclude that the difference must be related to a mistake in her numerical
computation.}.

The way we perform the numerical computation of the effective potential
using our approach is to separate the cases where the squared mass of the
Higgs boson (respectively the Goldstone boson) is positive or zero from that
where it is negative. In the first case, no problems appear in the numerical
evaluation of the effective potential, while we use expressions like
equation (\ref{potefft}) when the scalar masses squared become negative. The
rest of the expressions we use for the evaluation of the effective potential
are those given in ref. \cite{CARR}.

Figure 1 shows the effective potential at the critical temperature, for a
top-quark of mass 180 GeV, versus the value of $\phi $. The appearence of a
first order phase transition is shown. Notice that both results coincide
considerable well in a wide range, the differences appearing for values of $%
\phi $ bigger than the position of the non trivial minimum of the effective
potential. Indeed the critical temperature, defined when the minima of the
effective potential are degenerate, is the same when calculated through the
two different ways. Moreover, the ratio $\phi /T$ at the critical
temperature, which is the relevant parameter to be studied in order to
conclude something about the baryon asymmetry washout, is unaltered in both
calculations as Figure 2 shows.

Since both methods of evaluation of the effective potential differ by the
way how the Matsubara frequencies are taken into account (considering all
Matsubara frequencies in the first method and only the zero mode in our
proposal), the results show in a numerical way the irrelevance of those
terms contributed by the nonzero Matsubara frequencies.

Although our results do not differ significatively from those obtained
through the application of the self-consistent approach, our method has the
advantage of making the renormalization procedure conceptually clear,
avoiding the necessity of introducing temperature dependent counterterms.

\smallskip\

\noindent {\bf Acknowledgments}

\smallskip\

One of us (L.V.) wants to thank O. Espinosa for stimulating discussions in
earlier stages of this work and acknowledges the hospitality extended by the
people of the Departamento de F\'\i sica of Universidad Santa Mar\'\i a.
L.V. was supported by a FONDECYT postdoctoral fellowship, grant No. 3940005.
G.P. was supported in part by DICYT No. 049331PA and FONDECYT No. 1930067.

\newpage\

\noindent {\bf Figure Captions}

\smallskip\

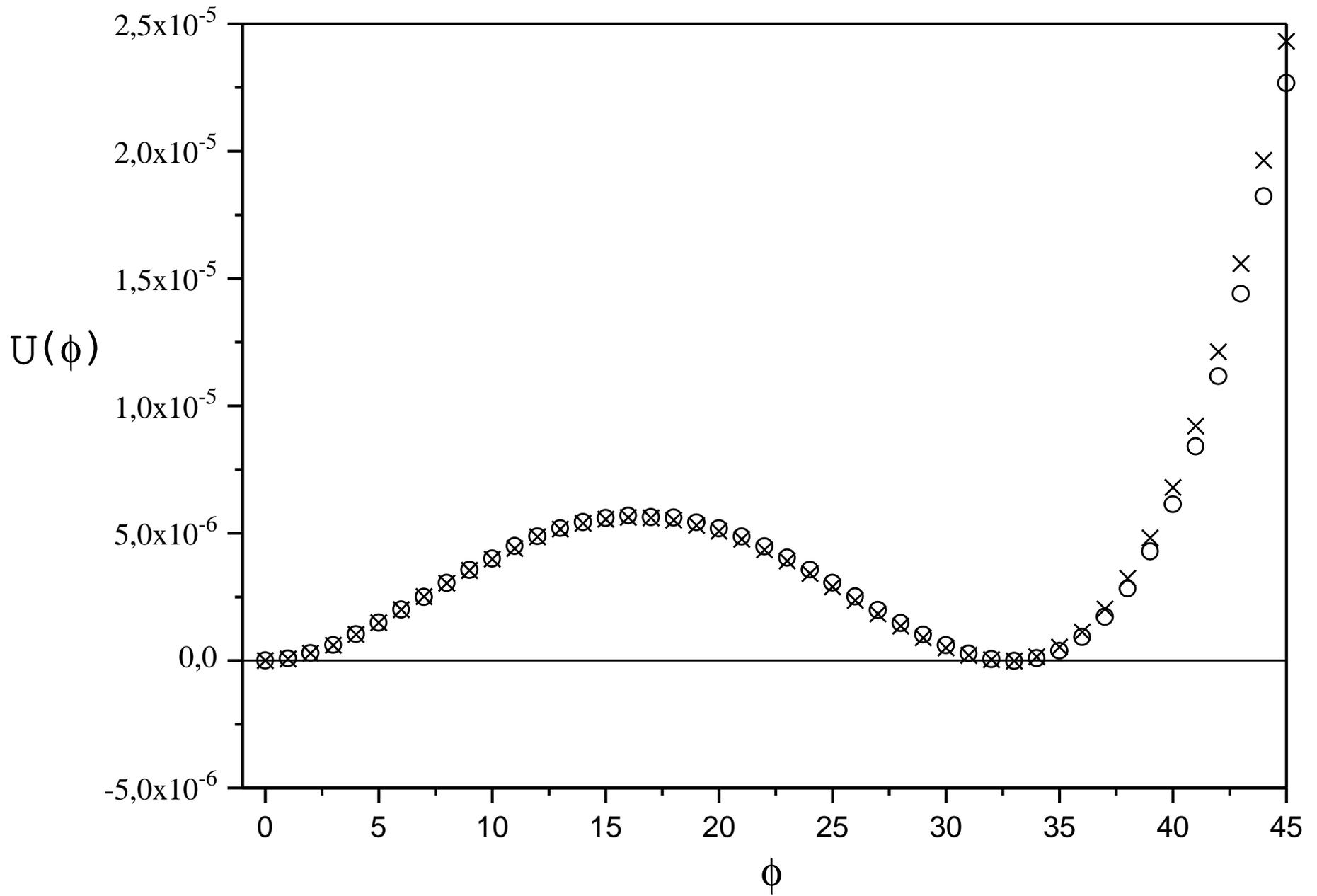
{\bf Figure 1}. The one-loop ring improved effective potential at finite
temperature U($\phi $) versus $\phi $ is shown, for a Higgs mass of 60 GeV
and a top-quark mass of 180 GeV. Circles represent the data from the
self-consistent method, while crosses are data from our proposal.

\smallskip\

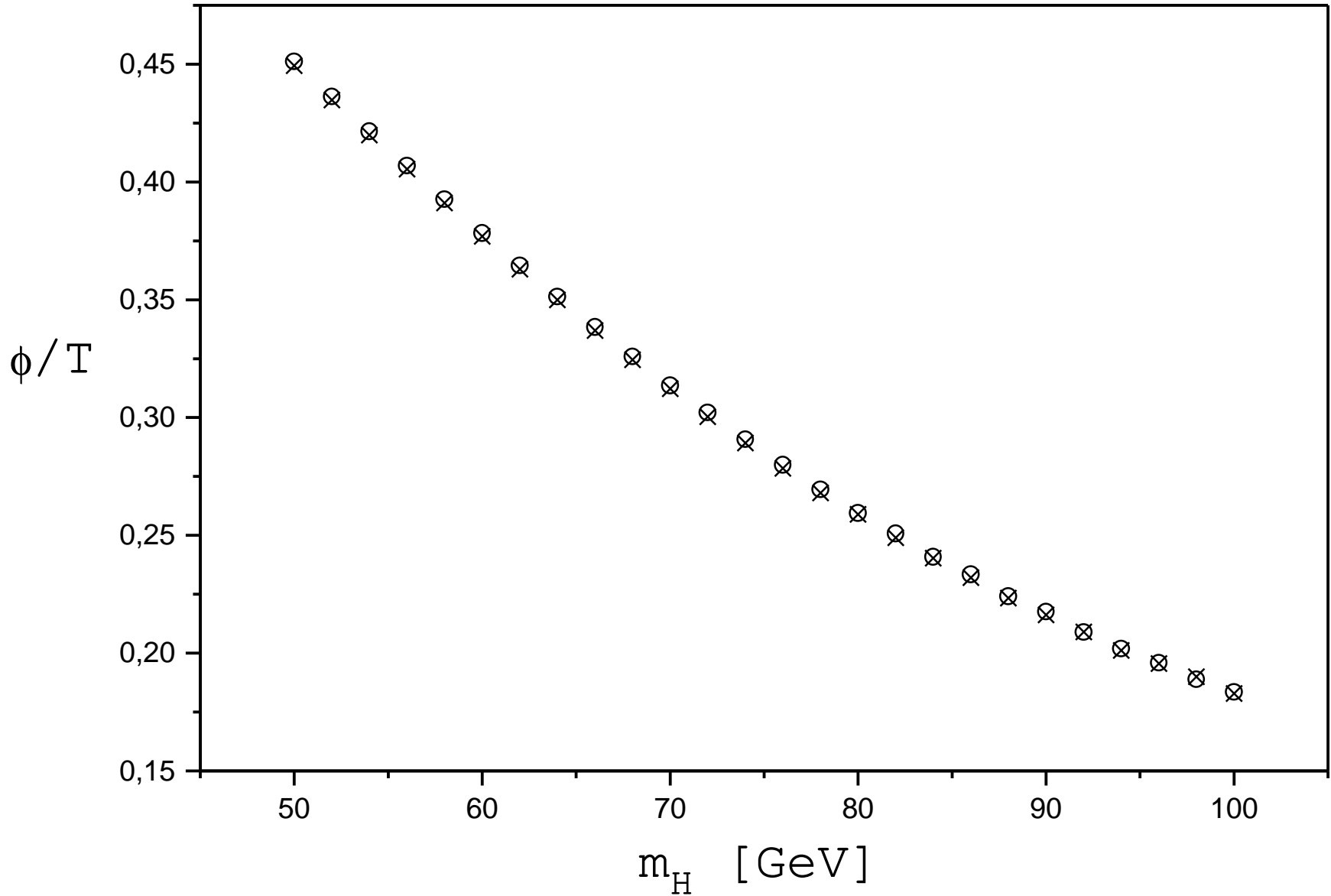
{\bf Figure 2}. The ratio $\phi /T$ for the same parameters as in figure 1
is plotted as a function of the Higgs mass. Notation is as in figure 1.

\end{document}